\documentclass[aip,amsmath,amssymb,preprint]{revtex4-1}
\usepackage{physics}
\usepackage{xcolor}
\usepackage{bm}
\usepackage{graphicx}

\newcommand{\mvec}[1]{\mathbf{#1}}

\makeatother

\begin{document}

\author{Mattia Bondanza}
\affiliation{Dipartimento di Chimica e Chimica Industriale, University of Pisa, via G. Moruzzi 13, 56124, Pisa, Italy}
\author{Tommaso Nottoli}
\affiliation{Dipartimento di Chimica e Chimica Industriale, University of Pisa, via G. Moruzzi 13, 56124, Pisa, Italy}
\author{Michele Nottoli}
\affiliation{Institute of Applied Analysis and Numerical Simulation, Universit\"at Stuttgart, Pfaffenwaldring 57, D-70569, Stuttgart, Germany}
\author{Lorenzo Cupellini}
\affiliation{Dipartimento di Chimica e Chimica Industriale, University of Pisa, via G. Moruzzi 13, 56124, Pisa, Italy}
\author{Filippo Lipparini}
\affiliation{Dipartimento di Chimica e Chimica Industriale, University of Pisa, via G. Moruzzi 13, 56124, Pisa, Italy}
\author{Benedetta Mennucci}
\email{benedetta.mennucci@unipi.it}
\affiliation{Dipartimento di Chimica e Chimica Industriale, University of Pisa, via G. Moruzzi 13, 56124, Pisa, Italy}

\title{The OpenMMPol Library for Polarizable QM/MM Calculations of Properties and Dynamics}

\begin{abstract}
We present a new library designed to provide a simple and straightforward way to implement QM/AMOEBA and other polarizable QM/MM methods based on induced point dipoles. The library, herein referred to as OpenMMPol, is free and open-sourced and is engineered to address the increasing demand for accurate and efficient QM/MM simulations.
OpenMMPol is specifically designed to allow polarizable QM/MM calculations of ground state energies and gradients, and excitation properties. Key features of OpenMMPol include a modular architecture facilitating extensibility, parallel computing capabilities for enhanced performance on modern cluster architectures, and a user-friendly interface for intuitive implementation and a simple and flexible structure for providing input data.
To show the capabilities offered by the library we present an interface with PySCF to perform QM/AMOEBA molecular dynamics, geometry optimization and excited state calculation based on (TD)DFT.
\end{abstract}

\maketitle

\section{Introduction}

Nowadays, hybrid models which combine a quantum chemical description with a classical but atomistic model based on molecular mechanics (MM) force fields have gained significant popularity for elucidating properties and processes of embedded molecules. QM/MM methods have shown to be extremely effective in describing not only homogeneous solvents but also much more complex environments characterized by extensive inhomogeneity and anisotropy.\cite{Warshel:2014fz,Senn:2009gk} Notably, these methods enable the explicit incorporation of electrostatic effects from the embedding environment into the QM description of electronic density, when utilizing the electrostatic embedding formulation.
The escalating popularity of QM/MM methods is further fueled by their ease of implementation in electronic structure codes. Specifically, incorporating the electrostatic influence of a group of MM atoms on the description of the QM charge density simply necessitates expanding the classical charges from the molecular nuclei to supplementary points situated outside the QM subsystem. This means that the the only part of the QM Hamiltonian which has to be changed is the one calculating the electrostatic potential of the nuclei.

Beyond the electrostatic embedding approach, QM/MM methods offer more comprehensive formulations. Specifically, leveraging a polarizable MM force field enables the integration of mutual polarization effects between the QM and MM subsystems. This can be achieved in different ways,
including fluctuating charges\cite{Rappe:1991ft,Steven_JCP_MMPolFQ,Lipparini:2011hu,Lipparini_JCTC_FQ2Der,Poier_JCTC_BCFQ,Cappelli_IJQC_FQOr}, Drude oscillators\cite{Boulanger:2012gx,Riahi_JCC_QMDrudeMD,Sahoo_Frontiers_CPMDDrude,Lu_JCTC_QMDrude}, or induced point dipoles (IPD) \cite{Warshel1976,Applequist_JACS_PPD,thole1981,Thompson:1995fc,Gao:1997de,vanduijnen98,Wang2011,curutchet_2009_electronicenergytransfer,Loco_JCTC_QMAMOEBA,Olsen_AQC_PE,Skylaris_JCTC_Tinktep,Skylaris_JCTC_2017,Skylaris_2017_XLBO,Skylaris_JCP_QMAMOEBA,Wu:2017jg}. 
In this contribution, we focus on the IPD approach.
Within this framework, each MM atom not only carries a fixed point charge but also possesses a typically isotropic polarizability. This property enables MM atoms to respond to applied electric fields by generating atomic induced dipoles. When these polarizable MM atoms embed a QM subsystem the field they experience arises from both the QM charge and the MM charges. The resulting induced dipoles introduce a new term in the QM Hamiltonian which, this time, depends on the QM charge density. As a result, the solution of the new Hamiltonian will correspond to QM and MM systems which have been mutually polarized. 
A popular example of the IPD formulation of polarizable embedding QM/MM is the one based on the AMOEBA force field.\cite{ponder2010amoeba} In the AMOEBA framework, each MM atom is additionally characterized by fixed multipoles extending up to quadrupoles.

The integration of mutual polarization alongside fixed-charge electrostatic effects is expected to enrich the description of environmental effects on molecular processes, especially when these processes involve multiple electronic states.\cite{Bondanza:2020he,Olsen_JCTC_PE} 
Nevertheless, while this enriched model offers increased comprehensiveness, it also introduces additional computational challenges. The appearance of a new QM-density-dependent term in the Hamiltonian in fact significantly increases the complexity with respect to the standard electrostatic embedding formulation. As a result, the implementation of polarizable QM/MM methods in electronic structure codes is far from being straightforward.
This complexity is further amplified when the focus extends beyond calculating the energy of the QM subsystem to encompass molecular properties, such as geometries, or dynamic processes. In all these cases in fact it is necessary to extend the polarizable QM/MM method to analytical gradients. 

IPD based polarizable QM/MM implementations are available in both commercial and open source quantum chemistry suite of programs. Our QM/MMPol and QM/AMOEBA machinery has been originally implemented into a development version of Gaussian.\cite{curutchet_2009_electronicenergytransfer,Loco_JCTC_QMAMOEBA,Lipparini2019} A similar approach based on the AMOEBA force field has also been implemented in the linear-scaling code OneTEP\cite{ONETEP}, which has been interfaced to Tinker\cite{Tinker} via the TINKTEP interface.\cite{Skylaris_JCTC_Tinktep}. IPD Polarizable QM/MM calculations are also available in deMon2K.\cite{Wu:2017jg,delalande2019}
However, the most commonly available model is the Polarizable Embedding model\cite{Olsen_JCTC_PE,Olsen_AQC_PE}, which has been implemented in two open-source libraries, PELib\cite{Olsen_JCTC_PE,Olsen_AQC_PE} and CPPE\cite{Scheurer_JCTC_CPPE,Scheurer21}, which have in turn been interfaced with  various commercial and opensource quantum chemistry packages. 

Our aim in developing OpenMMPol is to provide a flexible yet powerful platform to use the current state-of-the art atomistic polarizable force field in combination with virtually every QM method already implemented in any electronic structure code. Moreover, we would like to have, as much as possible, a clear and easy-to-understand implementation to allow using this framework as a development platform for new features. An important feature, and element of novelty, contained in OpenMMPol is that it provides all the features required to perform ab-initio, polarizable QM/MM MD simulations, as it includes all the bonded, non-bonded and QM/MM contributions to not only the energy, but also its gradients, including with respect the positions of the MM atoms. In other words, OpenMMPol contains a decade of developments carried out by our group towards general and efficient multiscale strategies to study molecular properties and dynamics.\cite{Nottoli_WIREs_QMAMOEBA}

\section{The OpenMMPol library}

OpenMMPol is built as a library that offers interfaces to the most popular language used in computational chemistry (C/C++, Python and Fortran). This design allows using the library in combination with most QM codes with a minimal impact on the build system and on the code itself. The interfaces exposed are as much as possible consistent between the different languages, with some adaptation to make their usage as natural as possible from each language. 

The code is written in modern Fortran, strictly adhering to the 2008ts standard and can therefore be compiled using different compilers suites; in particular GNU, Intel and NVidia compiler builds are tested using a continuous integration pipelines.
The code is compiled using the popular and flexible CMake build system and only requires few external libraries (namely BLAS\, LAPACK, OpenMP, OpenSSL, cJSON, and optionally HDF5 and PyBind11) which are  mature, stable and available on most Linux systems.

To interface a QM code with the OpenMMPol library, the host code should be written in Fortran, C/C++, or Python and it should be able to compute electrostatic potential and electric field one-electron integrals. To exploit all the features offered by OpenMMPol, and in particular to compute QM/AMOEBA energies and forces, field gradient and field hessian integrals should be available as well (\textit{vide infra}).
When higher/arbitrary order (\textit{e.g.} electric-field gradients) integrals are not readily available, they can be always computed by numerical differentiation of lower-order integrals that are available in most codes.

Internally, OpenMMPol is organized in several Fortran modules which are opaque to the final users. A single interface Fortran module (or an equivalent C library or Python module) exposes to the user all the functions, constants, and data structures required for interfacing an external QM code; as a convention, all the objects contained in the interface module have a name beginning with \texttt{ommp\_}. The code is documented using FORD, which allows one to generate HTML doc pages directly from the comments in the code. 

In the following sections we show the code's capabilities and the basics of the underlying implemented equations.

\subsection{Input and Data Structures}
The code is built around a main object called \texttt{ommp\_system} that represents the current status of the MM part of the system, together with all the parameters (electrostatics, dispersion-repulsion/van der Waals and bonded). Such object is a derived type in Fortran and is handled as a \texttt{void} pointer in C, and an actual object in Python. 
The library offers different options to initialize those objects from different file formats:
\begin{itemize}
    \item \texttt{.hdf5} this is a standard binary file format, which has been used to package all the information contained in the \texttt{ommp\_system} object for easy loading and saving. It has the advantage of being a binary file whose I/O can be easily parallelized. At the same time, since it is a widespread format in computational sciences, several applications and libraries allowing an easy inspection/modification of this format are available.
    \item \texttt{.xyz} + \texttt{.prm} according to Tinker manual specifications; this format is intended to be used for AMOEBA or other force fields implemented in both Tinker and OpenMMPol; currently is a good choice as it  allows using the Tinker infrastructure to prepare input files and then use it for QM/MM calculations with any QM code exploiting the OpenMMPol capabilities.
    \item \texttt{.mmp} a formatted file format containing only electrostatic and topological information about the system. It was developed and used in our group in earlier development phases of QM/MMPol methods and is implemented here for backward compatibility. As it cannot contain information about van der Waals (vdW) terms and bonded ones, it cannot be used for the calculation of the full gradients for QM/MM systems. This format is described in the Supplementary Material. 
\end{itemize}

To facilitate the interface with a QM code, a second object, called \texttt{ommp\_qm\_helper} is provided. This object represents the QM system within the library and is initialized using the coordinates and atomic numbers of the atoms of the QM subsystem. If required, a parameter file (in Tinker format) and atom types can also be assigned. This object is typically used to compute various electrostatic properties (potential, field, field gradient and Hessian) of the nuclei at the MM sites, the QM-MM component of the vdW interactions, and all the operations related to link atoms. Again, when the geometry of the QM system is altered (eg. during a geometry optimization, an MD simulation), the host code should also update the corresponding \texttt{ommp\_qm\_helper} object.

In order to make the modification of the input section of the host code as limited as possible, we have encapsulated all the OpenMMPol-related parameters in a single JSON file that is the only input required to initialize the library. In this file each possible option for a run (\textit{e.g.} input files, options for solving the linear system, required version of OpenMMPol, etc.) can be specified. A single interface call (\texttt{ommp\_smartinput}) does all the checks on both the JSON file and the file used to initialize the \texttt{ommp\_system} object; this interface routine sets all the relevant parameters and returns an \texttt{ommp\_system} object, and, if needed, an \texttt{ommp\_qm\_helper} object. This approach is beneficial when accessing all the functionality provided by the library (\textit{e.g.} QM/MM systems with link atoms) with a standard input file that can be used across different QM software packages interfaced to OpenMMPol. Moreover the use of a standard input format such as JSON allows either easy generation of inputs from scripts and modification by hand.

Even if lower-level initialization routines are available through the interface, we discourage their direct use as they are all accessible through \texttt{ommp\_smartinput} and a properly configured JSON file that also triggers several checks on the input that should otherwise be re-implemented in the host code. The parsing of JSON input relies on a small portion of code written in C99 exploiting the cJSON library, and interfaced to Fortran.

In the following, we briefly recapitulate the main equations implemented in OpenMMPol. We use the following notation conventions. Sums over cartesian components (e.g., $x,y,z$ for position vectors or dipoles, $xx, xy, xz, \ldots$ for quadrupoles and so one) are implicit and not shown. We use bold symbols for vectors and matrices that are collections of quantities, e.g., $\mvec{q}$ is a vector that collects all the point charges $q_i$ on MM atoms, or that are whole QM matrices, e.g., $\mvec{P}$ is the density matrix of the systems, the elements of which are $P_{\mu\nu}$.
\subsection{Electrostatics and polarization}
\label{sec:eeandpol}
OpenMMPol is primarily conceived to allow QM/MM calculations with polarizable MM models based on the induced point dipoles (IPD) formalism. Two distinct cases are addressed by the library: the AMBER polarizable force field \cite{Wang2011}, which closely resembles a standard AMBER FF with additional polarizabilities, and the AMOEBA force field \cite{ponder2010amoeba,Shi2013}. The latter differs from the former in two main aspects: first, the fixed part of electrostatics on each MM atom is represented with a multipolar expansion up to quadrupoles rather than just a point charge. The second difference, which is addressed more in detail in the last part of this Section, consists in a different definition for the electric field generated by the fixed multipoles used to \textit{induce} the dipoles, and the one used to compute the interaction with the induced dipoles.

The presence of fixed multipoles on MM atoms implicitly poses a further complication:
since dipoles and quadrupoles are vector quantities, in order to give a physically sound description of an atom's electric field, they should rotate together with the molecule. AMOEBA implements this by parametrizing dipoles and quadrupoles in an absolute reference frame and then rotating according to some parametrized rules to get them in the correct position for a particular geometry of the molecule.\cite{Lipparini_JCTC_Dipoles} In the following, unless explicitly stated, we always refer to the rotated multipoles.

OpenMMPol is able to compute all the energy and derivative terms related to electrostatics and polarization needed for a full QM/MM gradient calculation. Currently, we have implemented all the electrostatic and polarization terms using standard double loops that exhibit a $\mathcal O(N^2)$ scaling, including the solution to the polarization equations which is done using a preconditioned conjugate gradient iterative solver. We note that the scaling can be reduced to $\mathcal O(N)$ by using the fast multipoles method (FMM)\cite{Lipparini2019} to compute all such terms. A linear-scaling implementation will be distributed with the next release of OpenMMPol.

For the fixed electrostatic part, three interface functions are available:
\begin{itemize}
    \item \texttt{ommp\_get\_fixedelec\_energy} computes the interaction energy between the fixed charges (or multipoles, for AMOEBA) on each MM atom, according to
    \begin{equation}\label{eq:e_elec}
        \mathcal{E}^\mathrm{MM - elec} = \frac{1}{2}\sum_i\sum_{j\neq i}s_{ij}\sum_{L_1}\sum_{L_2}M_i^{L_1} T^{L_1L_2}_{ij} M_j^{L_2},
    \end{equation}
    where indices $i$ and $j$ refer to MM atoms whereas indices $L_x$ refer to angular momenta (0 for charges, 1 for dipoles, and so on). $s_{ij}$ is a scaling factor which accounts for the exclusion of neighbour electrostatics, $M_k^{L_x}$ is a Cartesian multipole centered on atom $k$ with angular momentum $L_x$. The components of the interaction matrix $T^{L_1L_2}_{ij}$ are defined as follows\cite{thole1981}, 
 \begin{equation}
        \label{eq:Coultens}
        [T^{L_1L_2}_{ij}]^{\alpha\beta\dots}_{\alpha^\prime\beta^\prime\dots} = \frac{\partial^{L_1}}{\partial r^\alpha_i\partial r^\beta_i\dots}\frac{\partial^{L_1}}{\partial r^{\alpha^\prime}_i\partial r^{\beta^\prime}_i\dots}\frac{1}{|r_i - r_j|}.
    \end{equation}

    We note that for the AMBER FF the angular momentum assumes only the value of zero, while for AMOEBA it ranges from zero to two, \textit{i.e.} up to quadrupoles.
    
    \item \texttt{ommp\_fixedelec\_geomgrad} computes the derivative of the interaction energy between fixed charges (or multipoles) with respect to the displacement of each MM atom. The implemented expression can be derived by differentiating Eq.~\eqref{eq:e_elec} with respect to the position of a MM atom $k$ and reads
    \begin{equation}\label{eq:derx_emm}
        \pdv{\mathcal{E}^\mathrm{MM - elec}}{r_{k,\rm MM}} = \sum_{j\neq k}s_{kj}\sum_{L_1}\sum_{L_2}M_k^{L_1} T^{L_1+1L_2}_{kj} M_j^{L_2}
    \end{equation}
    \item \texttt{ommp\_potential\_mm2ext} and \texttt{ommp\_field\_mm2ext} are used to compute the electrostatic potential and electric field respectively, generated by MM atoms at a set of arbitrary ``external" coordinates; \textit{i.e.},
    \begin{equation}
        \Phi_i^L = (-1)^{L}\sum_{j\neq i}\sum_{L_2}T^{LL_2}_{ij} M_j^{L_2},
    \end{equation}
    where $L=0$ for the potential, $L=1$ for the electric field, and so on.
\end{itemize}

In both AMBER and AMOEBA, each MM atom is further endowed with an isotropic polarizability $\alpha_i$. As a response to the electric field generated by both the QM and the MM parts of the system, an induced point dipole is generated on each atom.
The induced dipoles are obtained as the solution to the following linear system of equations

\begin{equation}\label{eq:ind_dip}
    \alpha_{i}^{-1} \mu_i + \sum_{j\neq i}s^{\mu}_{ij}\mathcal{T}^{11}_{ij}\mu_j = E_i^{\rm MM} + E_i^{\rm QM},
\end{equation}
whose right-hand-side contains the electric fields $E_i^{\rm MM}$ generated by the fixed multipoles and $E_i^{\rm QM}$ generated by the QM density. 
In Eq.~\ref{eq:ind_dip}, we have introduced a screening factor $s^{\mu}_{ij}$ that accounts for excluded interactions (\textit{i.e.}, for bonded atoms, or atoms belonging to the same polarization group depending on the force field) 
and the damped interaction matrix ($\mathcal{T}_{ij}$) \cite{thole1981}, defined as in Eq.~\ref{eq:Coultens} but replacing the standard Coulomb Kernel with $\lambda(r_{ij})/|r_i-r_j|$.
The damping function\cite{Sala2010,ponder2003force,thole1981,Wang2011} $\lambda$ ensures that, when the distance between two polarizable sites becomes small, the interaction energy does not diverge, as the kernel has a finite zero distance limit, avoiding thus the so-called \emph{polarization catastrophe}. We also underline that the screening factor used to compute $E_i^{\rm MM}$ in Eq.~\ref{eq:ind_dip} could differ from the one used in Eq.~\ref{eq:e_elec}.

Equations \ref{eq:ind_dip} can be collected in a matrix form, by introducing the $3n\times 3n$ matrix $\mathbf{T}$, whose $3\times 3$ blocks are defined as follows
\begin{equation}
    \label{eq:TMat}
    T_{ij} = \alpha^{-1}_i\delta_{ij} + (1-\delta_{ij}) \mvec{\mathcal{T}}^{11}_{ij},
\end{equation}
where $\delta$ is the Kronecker delta. The resulting linear system reads
\begin{equation}\label{eq:pol_eq}
    \mathbf{T}\bm{\mu} = \mvec{E}^{\rm MM} + \mvec{E}^{\rm QM} 
\end{equation}
and is solved by calling the OpenMMPol function \texttt{set\_external\_field}, which requires as input the QM electric field ($\mvec{E}^{\rm QM}$) calculated at the polarizable atom positions.
For a standard IDP-based force field, like AMBER, the polarization energy can thus be written as
\begin{equation}
    \label{eq:ene_pol}
    \mathcal{E}^\mathrm{Pol} = -\frac{1}{2}\sum_i\mvec{\mu}_i \left(E_i^{\rm MM} + E_i^{\rm QM}\right).
\end{equation}
In the case of AMOEBA force field, the problem is slightly more involved because the electric field that induces the dipoles is computed according to different screening rules with respect to the one used to compute the interaction energy.\cite{ponder2010amoeba,ponder2003force,ren2003polarizable} Across the code, and in the present paper, the first field is called \textit{direct} ($\mathbf{E}_d$) and the second one \textit{polarization} ($\mathbf{E}_p$). 

To deal with this specificity, the following polarization Lagrangian has been proposed\cite{lipparini2015polarizable,nottoli2020general}
\begin{equation}
    \label{eq:lagAMO}
    \mathcal{L}(\mu^d,\mu^p) = - \frac{1}{2} \sum_i \mu_i^d \left (E^{\rm MM,p}_i + E^{\rm QM}_i\right) + \sum_i \mu^p_i \left (  (\mvec{T} \bm{\mu}^d)_i - E^{\rm MM,d}_i - E^{\rm QM}_i\right )
\end{equation}
where a new set of induced dipoles, $\mu^p$, act as Lagrange multipliers to enforce that the IPDs $\mu^d$ satisfy the polarization equation. By imposing the stationarity of the Lagrangian with respect to both dipoles, we get two linear systems of equations of the type \ref{eq:pol_eq}, one referring to $\mu^p$ and the other to $\mu^d$ respectively. The polarization energy for AMOEBA thus reads:
\begin{equation}
    \label{eq:ene_amo}
    \mathcal{E}^\mathrm{Pol} = -\frac{1}{2}\sum_i\mu^d_i\left(E_i^{\rm MM,p} + E_i^{\rm QM}\right).
\end{equation}

\subsection{Nonelectrostatic terms}
Even if nonelectrostatic terms do not enter directly in the QM electronic Hamiltonian, 
they should not be overlooked, as they are essential to perform MD and geometry optimizations. OpenMMPol is able to compute all bonded and non-bonded non-electrostatic terms needed to perform calculations with AMOEBA and AMBER force fields.

Among the non-electrostatic terms a special attention is devoted to the implementation of the vdW interaction which is present in every pair of atoms and therefore it is $\mathcal O(N^2)$ scaling if implemented with a na\"ive double loop strategy. In our code, two slightly different expressions for this interaction are implemented: the 6-12 Lennard-Jones used in AMBER  \cite{Case2005}
\begin{equation}
    \mathcal E^\mathrm{VdW}_{i,j} = \epsilon_{i,j} \left[ \left(\frac{R_{ij}^0}{r_{ij}}\right)^{12} - 2 \left(\frac{R_{ij}^0}{r_{ij}}\right)^6\right],
\end{equation}
and the buffered 7-14 Lennard-Jones used for the AMOEBA forcefield \cite{ponder2010amoeba}:
\begin{equation}
    \mathcal E^\mathrm{VdW}_{ij} = \epsilon_{ij}\left(\frac{1+\delta}{\rho_{ij} + \delta}\right)^7 \left(\frac{1+\gamma}{\rho_{ij}^7 + \gamma} - 2\right),
\end{equation}
where $\rho_{ij} = r_{ij}/R^0_{ij}$, and $\delta$ and $\gamma$ are fixed values equal to $0.07$ and $0.12$, respectively.\cite{ren2003polarizable}
Independently of the model, VdW interactions decay very fast at long distances. Due to this characteristic, the calculation of VdW interaction energy can be approximated using a simple cutoff distance and calculated in a linear scaling computation time. To get the desired linear scaling we used an improved variation of the cell-lists algorithm \cite{Mattson1999}. To minimize the memory footprint, the cell lists themselves, i.e., the lists of particles contained in each cell in which the simulation space is divided, are saved in a compressed format similar to the Yale format for sparse matrices. The current implementation only allows to use a \textit{hard} cutoff without any smearing function. The exact value of this cutoff distance can be set in the JSON input file by the user. If no cutoff is specified here, the $\mathcal O(N^2)$ algorithm is used. 
The interface functions available for VdW are:
\begin{itemize}
    \item \texttt{ommp\_get\_vdw\_energy} and \texttt{ommp\_qm\_helper\_vdw\_energy} are used to compute the VdW energy of the MM part and of the QM-MM interaction.
    \item \texttt{ommp\_get\_vdw\_geomgrad} and \texttt{ommp\_qm\_helper\_vdw\_geomgrad} are used to compute the derivatives of the energy terms above with respect to the coordinates of the nuclei. While the first term has only contributions on MM atoms, the second one has contributions on both QM and MM atoms.
\end{itemize}

The bonded terms are very numerous, but they are generally inexpensive to compute, exhibit a naturally linear scaling and are very similar across force fields. To keep this section concise we will only resume the energy terms implemented in OpenMMPol with the corresponding functions.

\begin{table}[!ht]
    \centering
    \def\arraystretch{1.4}
    \begin{tabular}{c|c|c}
        Energy term & Potential function & Expression \\
        \hline
        bond stretching & \texttt{ommp\_get\_bond\_energy} & $k \Delta l^2 (1 + k_3\Delta l + k_4\Delta l^2)$\\
        Urey-Bradley angle & \texttt{ommp\_get\_urey\_energy} & $k \Delta l^2 (1 + k_3\Delta l + k_4\Delta l^2)$\\
        angle bending & \texttt{ommp\_get\_angle\_energy} & $k \Delta \theta ^2 (1 + \sum_{i=1}^4k_{i+2} \Delta \theta^i)$\\
        out-of-plane bending & \texttt{ommp\_get\_opb\_energy} & $k \Delta \theta ^2 (1 + \sum_{i=1}^4k_{i+2} \Delta \theta^i)$\\
        $\pi$-torsion & \texttt{ommp\_get\_pitors\_energy} & $k (1+ \cos(2\theta-\pi))$\\
        dihedral torsion & \texttt{ommp\_get\_torsion\_energy} & $\sum_{i=1,6} A_i \Big(1 + \cos(i\theta - \phi_i)\Big)$\\
        improper torsion & \texttt{ommp\_get\_imptorsion\_energy} & $\sum_{i=1,3} A_i \Big(1 + \cos(i\theta - \phi_i)\Big)$\\
        stretching-bending coupling & \texttt{ommp\_get\_strbnd\_energy} & -- \\
        stretching-torsion coupling & \texttt{ommp\_get\_strtor\_energy} & --\\
        bending-torsion coupling & \texttt{ommp\_get\_angtor\_energy} & -- \\
        torsion-torsion cmap & \texttt{ommp\_get\_tortor\_energy} & -- \\
    \end{tabular}
    \caption{Bonded energy terms implemented in OpenMMPol, name of functions needed to compute the energy terms and implemented potential functions. In general $\Delta l$ represents a length variation with respect to a reference lenght, $\Delta \theta$ an angle variation with respect to a reference angle, and $\theta$ are angles. All the other terms are parameters. }
    \label{tab:bonded_prm}
\end{table}
All the derivatives of those terms are also implemented in OpenMMPol and are readily available through interface functions that are completely analogous to the ones listed in Table \ref{tab:bonded_prm}.

\subsection{Link Atoms}
OpenMMPol implements link atoms to treat systems with a covalent bond across the QM and MM parts of the system. To use link atoms, in the JSON inputs the user should specify the atom type for every atom in the QM system, and the atoms involved in the bond to be cut. In practice the link atoms should be explicitly present in the QM part of the system; their coordinates are irrelevant as OpenMMPol automatically displace them in the correct position as soon as the calculation starts.
In the present implementation the link atom position is determined by:
\begin{equation}
    r_\mathrm{LA} = r_\mathrm{QM} + d_\mathrm{LA} \frac{r_\mathrm{QM} - r_\mathrm{MM}}{|r_\mathrm{QM} - r_\mathrm{MM}|}.
\end{equation}
The parameter $d_\mathrm{LA}$ has a default value of 1.1~{\AA} which can be modified through the JSON input file.
The interactions in the link atom region are treated using the following rules\cite{Nottoli2020}:
\begin{itemize}
    \item the stretching term, the bending terms and the torsion terms involving the QM and the MM atoms are added according to FF rules, 
    \item QM-MM VdW interactions are screened according to FF rules and neglected for the link atom,
    \item Fixed electrostatic multipoles and polarizabilities are removed from the MM atom involved in the bond with the QM atom, as well as on every atom within two bonds from it (if needed the distance can be increased to an arbitrary number of bonds through JSON input),
    \item Charges are removed from the linked MM atom and its neighbours and they are redistributed evenly on the first unaffected neighbour shell of MM atoms, 
\end{itemize}
Those rules, according to our experience, allow performing stable MD simulations and optimizations with minimal user effort. It should be noted that, while for a simulation without link atoms the only required parameters for QM atoms are the VdW ones, here also bonded parameters are required. 


\section{Coupling to QM codes}
In this section we show how to couple the OpenMMPol library to an existing electronic structure code. As an example, we use the popular QM software PySCF.\cite{Sun2017,Sun2020} 
Alongside with the presentation of the interface we also provide some guidelines that can be  helpful when interfacing a new software to the library.
In particular, we start by describing how to couple OpenMMPol to a Hartree-Fock Self-Consistent Field (HF-SCF) code. To keep the notation simple, we limit the discussion to the HF case, but the generalization to Kohn-Sham density functional theory (DFT) is straightforward. 

\subsection{SCF Methods}
Introducing an atomic orbital (AO) basis, the SCF generalized eigenvalue problem to be solved is
\begin{equation}
    \label{eq:HF_eigen}
    \mvec{F}\mvec{C} = \mvec{S}\mvec{C}\mvec{\varepsilon},
\end{equation}
where $\mvec{F}$ is the Fock matrix, $\mvec{C}$ are the molecular orbital (MO) coefficients, $\mvec{\varepsilon}$ is a diagonal matrix holding MO energies, and $\mvec{S}$ is the AO overlap matrix. 
The Fock matrix contains new terms due to the interaction between the QM subsystem and the MM fixed multipoles and induced dipoles:
\begin{equation}
\label{eq:Fock}
     F_{\mu\nu} = h_{\mu\nu} + h_{\mu\nu}^\mathrm{MMPol}+ 
     G_{\mu\nu}(\mvec{P}) + G_{\mu\nu}^\mathrm{MMPol}(\mvec{P}),
\end{equation}
 where the density-dependent term $G_{\mu\nu}^\mathrm{MMPol}(\mvec{P})$ describes the interaction of QM density with induced dipoles, and the density-independent $h_{\mu\nu}^\mathrm{MMPol}$ for the one with fixed multipoles. 
 These two terms read:
\begin{align}
    &h_{\mu\nu}^\mathrm{MMPol} = \sum^\mathrm{MM}_i\sum_{L_1} (-1)^{L_1} M_i^{L_1} \langle\chi_\mu|\hat{t}^{L_1}_i|\chi_\nu\rangle,\label{eq:h_mmpol}\\
    &G_{\mu\nu}^\mathrm{MMPol}(\mvec{P}) = -\sum^\mathrm{MM}_i \mu_i(\mvec{P}) \langle\chi_\mu|\hat{t}^1_i|\chi_\nu\rangle, \label{eq:veff_mmpol}
\end{align}
the operator $\hat{t}^L_{i}$ being 
\begin{equation}
    \hat{t}^L_{i} = \frac{\partial^L}{\partial r^\alpha_i\partial r^\beta_i\dots}\frac{1}{|r-r_i|};
\end{equation}
we remind here that $L_1 = 0$ for AMBER, while  $L_1 = 0,1,2$ for AMOEBA which contains up to fixed quadrupoles. When AMOEBA is considered, attention should also be payed to the induced dipoles used in Eq.~\eqref{eq:veff_mmpol} which are the average of $\mu^d$ and $\mu^p$. 

IPDs in Eq.~\eqref{eq:veff_mmpol} are computed according to Eq.~\eqref{eq:pol_eq}, and therefore they depend on the electric field generated by the QM electronic density, here represented by the density matrix $\mvec{P}$. This shows the mutual coupling between the QM and MM subsystems; in HF (and DFT) this mutual dependency can be easily addressed within the SCF iterative framework.
At each SCF cycle the electric field generated by the current electronic density at polarizable sites is computed, the linear system (\textit{cf.} Eq.~\eqref{eq:pol_eq}) is solved, and finally the interaction matrix in the AO basis (that is, last term of Eq.~\eqref{eq:veff_mmpol}) is computed and added to the Fock matrix. All the other terms do not present any further complication as they are exactly the same as in standard electrostatic embedding coupling.

\begin{figure}[!ht]
     \centering
     \includegraphics[scale=0.5]{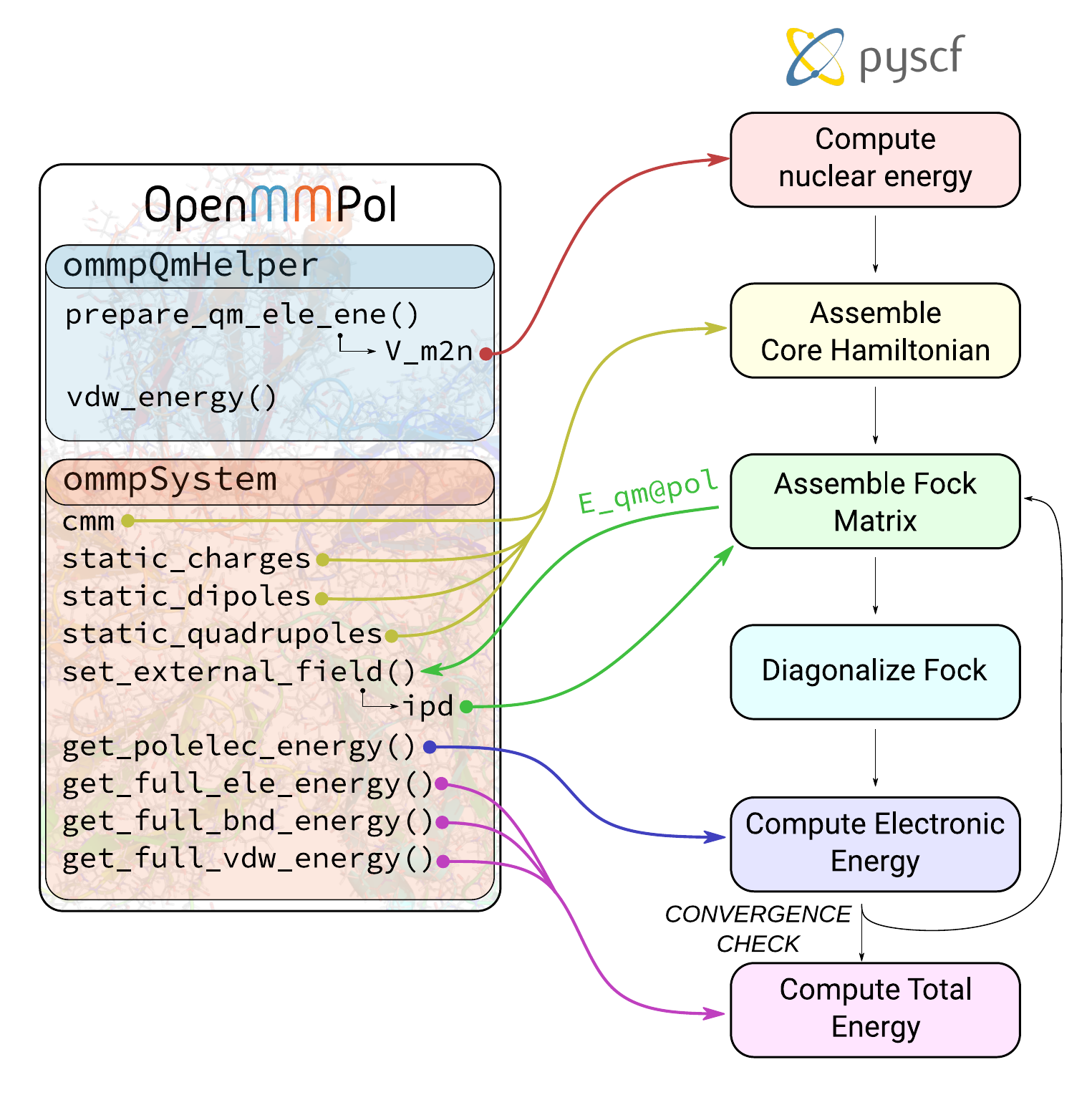}
     \caption{Graphical representation of the interface between OpenMMPol and PySCF. Note that the diagonalization of Fock matrix and the machinery of the self-consistent field is left untouched except for the extra terms added to the electronic energy and to the Fock matrix.}
     \label{fig:SCF_scheme}
 \end{figure}

To get the interface with the QM code working, it is necessary to alter the definition of the Fock matrix according to  Eqs.~\eqref{eq:Fock}--\eqref{eq:veff_mmpol}. To maintain the logic of standard SCF, and to avoid double calculations, it is convenient to spot several regions of the code where the modification will be introduced (see Fig.~\ref{fig:SCF_scheme}). 
When the QM code computes the repulsion between nuclei, it should add the interaction between nuclei and MM charges (or the fixed multipoles, for AMOEBA) computed as dot product between \texttt{ommp\_qm\_helper\%V\_m2n} and the nuclear charges. Here other density-independent energy terms can be computed and added: bonded interactions, vdW interactions, MM electrostatic interactions except the ones with the induced dipoles. In PySCF we overloaded the function \texttt{energy\_nuc}. Moreover, when the QM code computes the ``core Hamiltonian'', the effect of the MM charges (or the fixed multipoles, for AMOEBA) should be added (Eq.~\eqref{eq:h_mmpol}). Depending on the size of the system it could be convenient to compute the resulting atomic integrals on the fly or to store them in memory. In PySCF we overloaded the function \texttt{get\_hcore}.
Finally, when the QM code assembles the Fock matrix at each SCF iteration, it should also add the non-linear component that characterizes the polarizable embedding (Eq.~\eqref{eq:veff_mmpol}); to do so several steps are required:
    \begin{enumerate}
        \item the electric field $\mvec{E}^{\rm QM}$ generated by the current QM electron density at the MM sites is computed;
        \item the electric field is fed to OpenMMPol via \texttt{set\_external\_field}; note that this field should also account for the nuclear charges of the QM subsystem. This triggers the solution of the linear system inside OpenMMPol.
        \item the induced dipoles are retrieved from the MMPol Object via \texttt{ommp\_system\%ipd} and used to compute the polarization contribution to the Fock matrix (Eq.~\ref{eq:veff_mmpol}) and to the energy. 
    \end{enumerate}
    In PySCF we overloaded the function \texttt{get\_veff}.

The overall result is that at each SCF cycle an updated linear system is solved, and new IPDs are computed. In our experience, this procedure does not deteriorate the convergence of the SCF algorithm if a tight enough convergence of the polarization equations is enforced. We also notice that during each SCF cycle the electric field integrals are used twice (once for computing the density's electric field and once to compute $G_{\mu\nu}^\mathrm{MMPol}$); it is therefore convenient, unless very fast integrals routines are available, to store those integrals in memory during the whole SCF procedure.

As a general remark, when interfacing with a new code it is generally a good idea to proceed step-by-step going from trivial cases (\textit{e.g.} simple electrostatic embedding without polarizabilities) to intermediate ones (\textit{e.g.} MMPol) to reach the most difficult ones (\textit{e.g.} AMOEBA). When trying to implement some method for which there is no reference available, it could be a good idea to compare the analytical Fock matrix, with a numerical one obtained computing the numerical derivative of the energy with respect to a small variation of the density matrix. 

As a result of the SCF cycle, the QM/MM interaction energy reads
\begin{equation}
    \label{eq:QMMM_ene}
    \mathcal{E}^{\rm QM/MM} = \sum_{\mu\nu}P_{\mu\nu}h_{\mu\nu}^\mathrm{MMPol} + \frac{1}{2}\sum_{\mu\nu}P_{\mu\nu}G_{\mu\nu}^\mathrm{MMPol}(\mvec{P}) + \sum_i^\mathrm{QM}\sum_j^\mathrm{MM} Z_i \left (\sum_{L_1} T^{0 L_1}_{i,j} M_i^{L_1} - \frac{1}{2} T^{01}_{ij}\mu_i(\mvec{P})\right ),
\end{equation}
which, together with the QM and purely MM energy, gives the total energy. Note that, as the QM/MM vdW contribution does not depend on the electronic density, we have implicitly included them in the MM energy.


\subsection{Forces}
Gradients of the energy with respect to the atomic coordinates are a fundamental tool as they enable molecular dynamics (MD) simulations and geometry optimizations. 
Analytical expressions for gradients are easily derived from the energy; starting from gradients with respect to MM coordinates:
\begin{equation}
    \label{eq:grd_wrt_mm}
    \frac{\partial \mathcal{E}}{\partial r_{k,\mathrm{MM}}} = 
    \frac{\partial \mathcal{E}^\mathrm{MM}}{\partial r_{k,\mathrm{MM}}} +
    \frac{\partial \mathcal{E}^\mathrm{QM}}{\partial r_{k,\mathrm{MM}}} +
    \frac{\partial \mathcal{E}^\mathrm{QM/MM}}{\partial r_{k,\mathrm{MM}}}
\end{equation}
where the $\frac{\partial}{\partial r_{k,\mathrm{MM}}}$ denotes the partial derivative with respect to a specific coordinate of the $k$-th MM atom. The first term in Eq.~\eqref{eq:grd_wrt_mm} corresponds to the derivative of the MM energy (including VdW interaction between QM and MM atoms) and it can be easily computed as the MM energy function is in a closed analytical form (see for example Eq.\eqref{eq:derx_emm} for the electrostatic component). The second term vanishes and the last term can be obtained by differentiating the QM/MM interaction energy Eq.\eqref{eq:QMMM_ene}. 
The fixed electrostatic and polarization contributions are given by:
\begin{align}\label{eq:der_e_qmmm}
\frac{\partial\mathcal{E}^\mathrm{QM/MM}}{\partial r_{k,\mathrm{MM}}} = 
&\sum_L M_k^L \sum_{\mu\nu}P_{\mu\nu}\bra{\chi_\mu}\hat{t}_k^{L+1}\ket{\chi_\nu} \\
& - \mu_k \cdot \sum_{\mu\nu}P_{\mu\nu}\bra{\chi_\mu}\hat{t}_k^{2}\ket{\chi_\nu} \notag \\
&+ 
\sum_i^\mathrm{QM} Z_i \left ( \sum_{L} T^{1 L}_{i,k} M_k^{L} - T^{11}_{i,k} \mu_k \right );\notag
\end{align}
as noticed for Eq.~\ref{eq:veff_mmpol}, when AMOEBA is considered, $\mu_k$ is the average of $\mu^d$ and $\mu^p$.

Gradients with respect to QM coordinates are computed similarly:
\begin{equation}
    \label{eq:grd_wrt_qm}
    \frac{\partial \mathcal{E}}{\partial r_\mathrm{QM}} = 
    \frac{\partial \mathcal{E}^\mathrm{MM}}{\partial r_\mathrm{QM}} +
    \frac{\partial \mathcal{E}^\mathrm{QM}}{\partial r_\mathrm{QM}} +
    \frac{\partial \mathcal{E}^\mathrm{QM/MM}}{\partial r_\mathrm{QM}}
\end{equation}
where the first term reduces to the component on QM atoms of QM-MM vdW interaction, and the second is the standard geometric derivative of the SCF energy. 
The last term is again computed explicitly from the energy Eq.~\ref{eq:QMMM_ene}:

\begin{align}
     \label{eq:grd_wrt_qm_eqmmm}
     \frac{\partial\mathcal{E}^\mathrm{QM/MM}}{\partial r_{i,\mathrm{QM}}} =   &\sum_k \sum_L M_k^L  \sum_{\mu,\nu} P_{\mu \nu}  \frac{\partial}{\partial r_{i,\mathrm{QM}}} \langle\chi_\mu | \hat{t}_k^L | \chi_\nu\rangle \nonumber \\ 
     &- \sum_k \mu_k \sum_{\mu,\nu} P_{\mu \nu} \frac{\partial}{\partial r_{i,\mathrm{QM}}} \langle \chi_\mu |\hat{t}_k^L | \chi_\nu \rangle\nonumber \\
     &+ \sum_k^\mathrm{MM}Z_i \left ( \sum_{L} T^{1 L}_{i,k} M_k^{L} - T^{11}_{i,k} \mu_k \right )
\end{align}

It is useful to underline that all the quantities needed to compute the analytical MMPol-SCF gradients are one-electron integrals contracted with the QM density matrix and the proper electrostatic feature. Moreover, all those integrals (up to arbitrary order) could be efficiently evaluated using PRISM \cite{johnson1993computing} or similar algorithms as three-center two-electron integrals (see Section I of the Supplementary Material for further details).

From a practical point of view, when an SCF code is interfaced with OpenMMPol, the forces on QM atoms should be modified in order to include the terms in Eq.~\ref{eq:grd_wrt_qm_eqmmm}; the calculation of first two terms should be inside the QM code, as they are simple contraction of AO integrals that are computed with highly efficient generic matrix-matrix library function (eg. \texttt{dgemm}); the last term can be computed exploiting the \texttt{qm\_helper} object, which is able to compute the required derivatives of QM nuclei electrostatic potential.
Two density independent terms should also be added: the one corresponding to the QM-MM vdW force, and the one due to link-atoms; both of them are easily computed with OpenMMpol calling \texttt{ommp\_qm\_helper\_vdw\_geomgrad} and \texttt{ommp\_qm\_helper\_link\_atom\_geomgrad} rispectively.

The forces on MM atoms should also be computed inside the host code, following Eq.~\ref{eq:grd_wrt_mm}. All the purely MM components are easily computed inside the library calling the routine \texttt{ommp\_full\_geomgrad} (which is in turn a shorthand for calling \texttt{ommp\_full\_bnd\_geomgrad}, \texttt{ommp\_polelec\_geomgrad}, \texttt{ommp\_fixedelec\_geomgrad} and \texttt{ommp\_vdw\_geomgrad}); the MM components of QM/MM vdW and link-atoms forces are computed by the same routine named for the corresponding QM components. 
All the terms from the QM/MM electrostatic coupling in Eq.~\ref{eq:der_e_qmmm} are computed inside the QM code in a way that closely resembles the one described for Eq.~\ref{eq:grd_wrt_qm_eqmmm}.
If AMOEBA is used a last additional term should be added as a consequence of multipole rotation described in Section \ref{sec:eeandpol}. This term is often called torque force across the literature and it basically describes the force needed to reorient an atom with an anisotropic electrostatic potential inside the constant field generated by the other atoms and the QM density. The derivation of this terms can be found in ref. \citenum{Lipparini_JCTC_Dipoles}. Here it suffices to say that it can be computed by the library, using the function \texttt{ommp\_rotation\_geomgrad} which requires the QM density's electric field and electric field gradient.

\subsection{Excited states}
In the context of time-dependent (TD) HF and DFT, transition properties such as excitation energies and transition moments are customarily obtained as poles and residues of the linear-response (LR) function, respectively.\cite{Dreuw2005Review} In practice, a generalized eigenvalue system of equations, known as Casida equations, needs to be solved. When real-valued orbitals are used, it assumes the following form
\begin{equation}\label{eq:casida}
    \begin{pmatrix}
        \mathbf{\Tilde{A}} & \mathbf{\Tilde{B}} \\
        \mathbf{\Tilde{B}} & \mathbf{\Tilde{A}}
    \end{pmatrix}
    \begin{pmatrix}
        \mathbf{X}_K\\
        \mathbf{Y}_K
    \end{pmatrix}
    =
    \omega_K
    \begin{pmatrix}
        \mathbf{1} & \mathbf{0}\\
        \mathbf{0} & -\mathbf{1}
    \end{pmatrix}
    \begin{pmatrix}
        \mathbf{X}_K\\
        \mathbf{Y}_K
    \end{pmatrix},
\end{equation}
$\omega_K$ being the $K$-th transition energy, and $\mathbf{X}_K$ and $\mathbf{Y}_K$ transition densities. The structure of the $\mathbf{\Tilde{A}}$ and  $\mathbf{\Tilde{B}}$ orbital rotation Hessian matrices is modified with respect to their usual form by the addition of a polarization term $\mathcal{V}^{\rm pol}_{ai,bj}$ directly influencing the system's response
\begin{equation}
    \label{eq:vpol_td}
    \mathcal{V}^{\rm pol}_{ai,bj} = -\sum_k^{N_{\rm pol}}E_{ai}(\mvec{r}_k)\mu^T_p(\phi_b\phi_j),
\end{equation}
whose expression is the second derivative of the polarization Lagrangian with respect to the elements of the electronic density. 
$\bm{\mu}^T$ is the dipole induced by the electric field generated by the transition density element $\phi_b(\mvec{r}) \phi_j(\mvec{r}) $, and $E_{ai}(\mvec{r}_k)$ is the field computed at the $k$-th polarizable site arising from $\phi_a(\mvec{r}) \phi_i(\mvec{r})$.
Also, we note that the molecular orbitals $\phi(\mvec{r})$ and their energies stem from the solution of the modified SCF problem described in the previous section. The generalized eigenvalue problem and the polarization equation for $\bm{\mu}^T$ are solved iteratively in a self-consistent scheme.

This LR approach accounts only for a component of the response of the polarizable embedding, namely the one represented by the IDPs generated by the QM transition density. A second component instead should be computed by considering the IDPs generated by the change in the QM state density. In the literature, the former component is generally indicated as LR and the latter one as state-specific (SS).
To recover the SS component in a LR formulations such as TDDFT, various models have been proposed. \cite{caricato2006formation,improta2006state,zeng2015analytic} Here, we present the one that is more commonly used, e.g. the so-called corrected-LR (cLR)\cite{caricato2006formation} approximation. Within this framework, a correction to the excitation energy is computed according to the following expression:
\begin{equation}\label{eq:cLR}
    \Delta\omega_K^{\rm cLR} =  - \frac{1}{2}\sum_k^{N_{\rm pol}}\mu_k(\mvec{P}^\Delta_K)E_k(\mvec{P}^\Delta_K),
\end{equation}
where $\mvec{P}^\Delta_K$ is the change in the density matrix from the ground state to the excited state $K$. The latter requires solving the so called Z-vector equations.\cite{Caillie2000,Furche:2002kj,Scalmani:2006it}
Finally, the cLR correction \eqref{eq:cLR} can be directly added to the solution to the Eq.~\eqref{eq:casida}; this protocol was recently called cLR$^2$.\cite{guido2021simple}

Practically, the extension to the TDHF/TDDFT level of theory in PySCF only required overloading the  function \texttt{gen\_response} by adding the contribution given by Eq.~\eqref{eq:vpol_td}. More specifically, at each iteration of the Davidson solver the electric field stemming from the transition density matrix is computed and the polarization equation is solved. Note that the nuclear contribution to the electric field should be discarded: this can be specified by setting to \texttt{True} the optional argument \texttt{exclude\_nuclei} of the function \texttt{set\_external\_field}.


\section{Results}

In this Section we report an in-depth analysis of the performance of the library and of the interface code that allows to perform SCF calculuation using PySCF. All the tests have been performed on a single compute node equipped with two AMD EPYC 7282 processors (32 cores with multi-threading enabled) and 250~Gb of RAM. OpenMMPol was compiled with the Intel compiler v. 2023.1 using the \texttt{RELEASE} flags provided in the build system. PySCF was compiled with the GNU 12 compiler with the default option of the build system; both the codes are linked against the Intel MKL linear algebra libraries.

To test the performance of the library itself, we used a simple application that performs an initialization from a Tinker \texttt{.xyz} and \texttt{.prm} parameter file, computes all the electrostatic/polarization, bonded, and vdW contributions to the energy and the corresponding geometrical gradients reporting timings for each operation. 

\begin{figure}[!ht]
     \centering
     \includegraphics[scale=0.4]{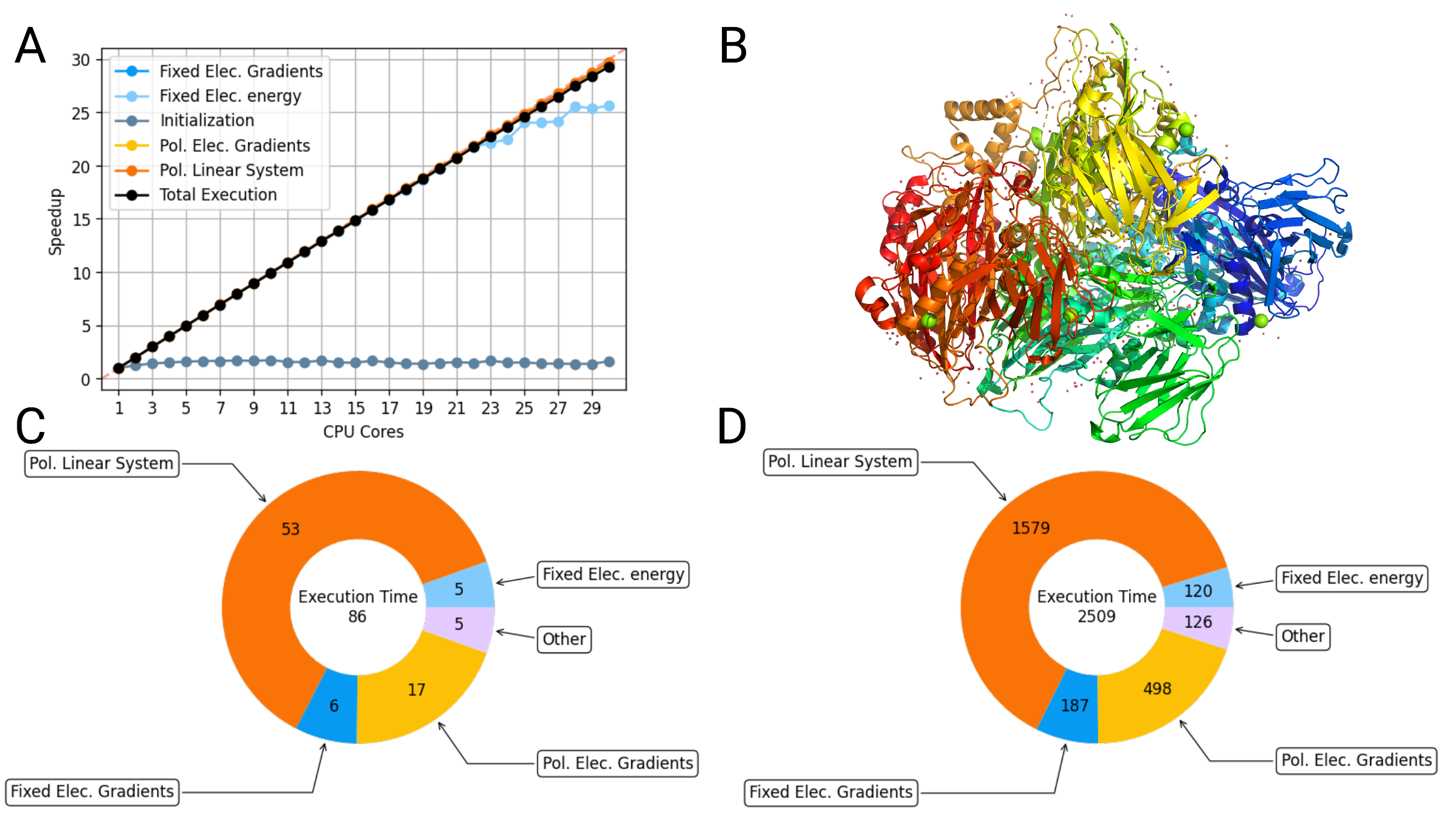}
     \caption{Results of parallelization performance tests performed on 1OHF. (A) Speed-up vs number of processor curve; the ideal speedup is indicated with the dashed red line, curves of the major contribution are depicted according to the legend and the total scaling is shown in black. (B) Representation of the simulated protein assembly (1OHF). (C,D) Major CPU time contribution of the calculation when running with 1 CPU (D) and with 30 CPUs (B).}
     \label{fig:scalingCPU}
 \end{figure}

First, we benchmarked the performance of our parallel implementation; to do this we performed the calculation on the  \textit{Nudaurelia capensis} $\omega$ virus capsid protein\cite{Helgstrand2004} (PDB: 1OHF, see Figure~\ref{fig:scalingCPU}B) composed of 35~695 polarizable atoms with the AMOEBA force field, using from 1 to 30 processors. Results are shown in Figure \ref{fig:scalingCPU}.

\begin{figure}[!ht]
     \centering
     \includegraphics[scale=0.4]{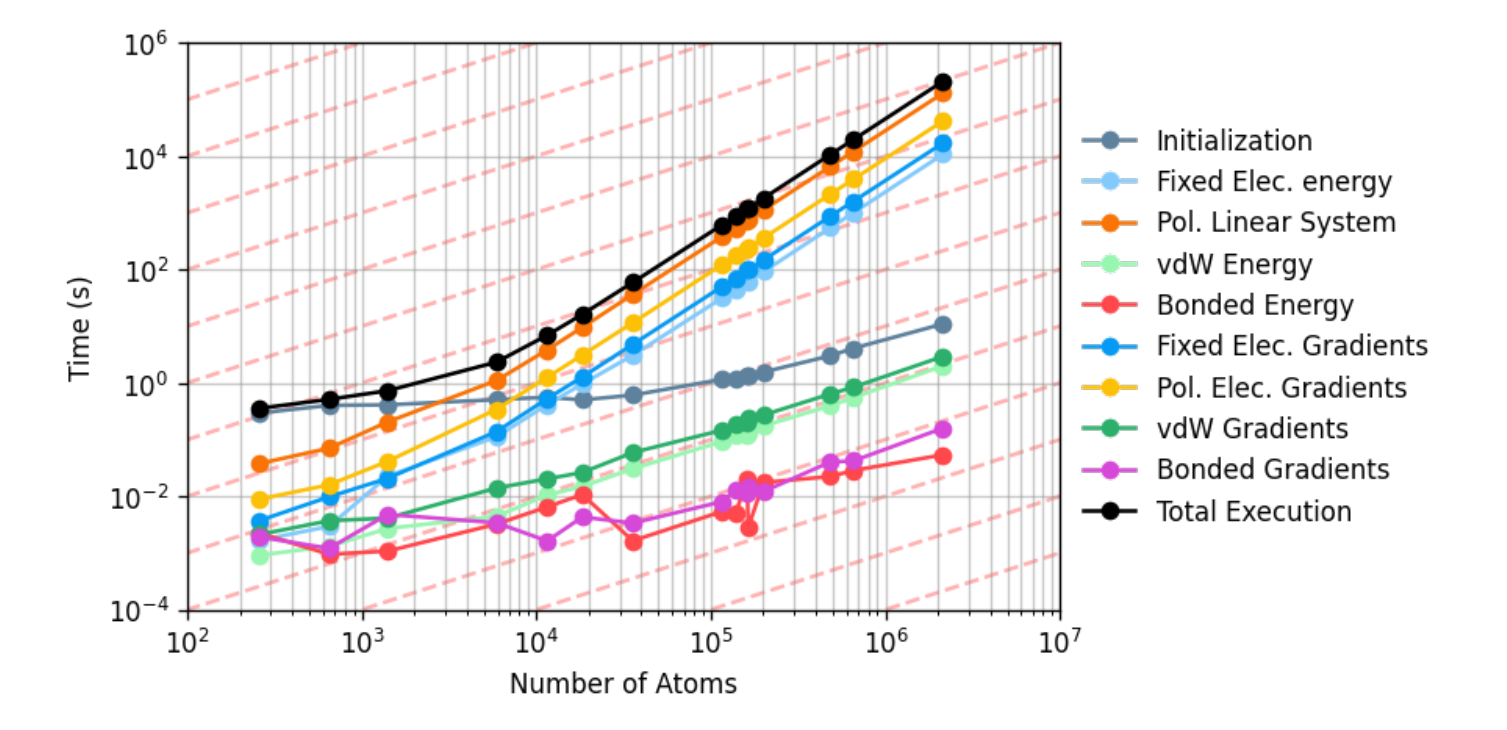}
     \caption{Time spent in different parts of the code for different system's size. We have always used protein systems, computing vdW interactions with a 12~\AA~cutoff and initializing the system from Tinker's \texttt{.xyz} and \texttt{.prm} files. Results are shown in a log/log graph and ideal linear scaling trends at each order of magnitude are reported as red dashed lines.}
     \label{fig:scalingSize}
 \end{figure}

 It can be seen that our library show a very nice overall scaling, almost perfectly adhering to the ideal trend. The only part of the code that does not have a good scaling is the initialization. After investigation, it is clear that for systems with less than $\sim$10k atoms, the ``constant'' part of the initialization (\textit{e.g.} parameter file parsing) is the most important one (see Fig. \ref{fig:scalingSize}); unfortunately this part is not easily parallelizable, and the situation cannot be significantly improved without a complete strategy change. 
 For systems with more than 100k atoms, the system-dependent part (which can be to some extent parallelized, see Fig.~\ref{fig:scalingSize}) becomes predominant, and the computational cost exhibits a nice linear scaling trend in the system size and a better scaling with respect to the number of CPUs. 
 It should be anyway underlined that the initialization for systems of less than 100k atoms is pretty fast (requiring less than 1~s on our test node), and it is done only once per simulation (while the polarization routine is repeated every SCF cycle and all other terms are computed at each geometry change). In this sense the situation depicted in Fig.~\ref{fig:scalingCPU}C is a worst-case scenario. Since all the other components have an almost ideal scaling, analyzing either panel B or C of Fig.~\ref{fig:scalingCPU} allows us to draw the same conclusions. For the system under analysis, all the time-consuming parts of the calculations are electrostatic-related, and the solution of the linear system in particular accounts for almost 2/3 of the whole execution time. This is not surprising as in our code electrostatics is computed with a na\"ive double loop strategy, which proves here its shortcomings.

 We have also performed a similar test, using a whole compute node, to compute energy and gradients of systems of different sizes, again without any coupling with QM codes.
Results are reported in Fig.~\ref{fig:scalingSize}. It is clear from these data that every part of the code, excluding electrostatics, has the ideal linear-trend scaling. Electrostatic contributions instead show, as expected, a $\mathcal{O}(N^2)$ scaling. As a consequence for every system with more than 1k atoms, the electrostatics becomes almost the only relevant part of the calculation, and all the other contributions are negligible with respect to it. This sub-optimal scaling also sets the current upper bound to the size of tractable systems on an average compute node between 10k and 100k polarizable atoms. This problem can however be tackled using a fast multipoles method algorithm to obtain a linear-scaling computational cost in the system's size for any required numerical accuracy. 

\begin{figure}[!ht]
     \centering
     \includegraphics[scale=0.4]{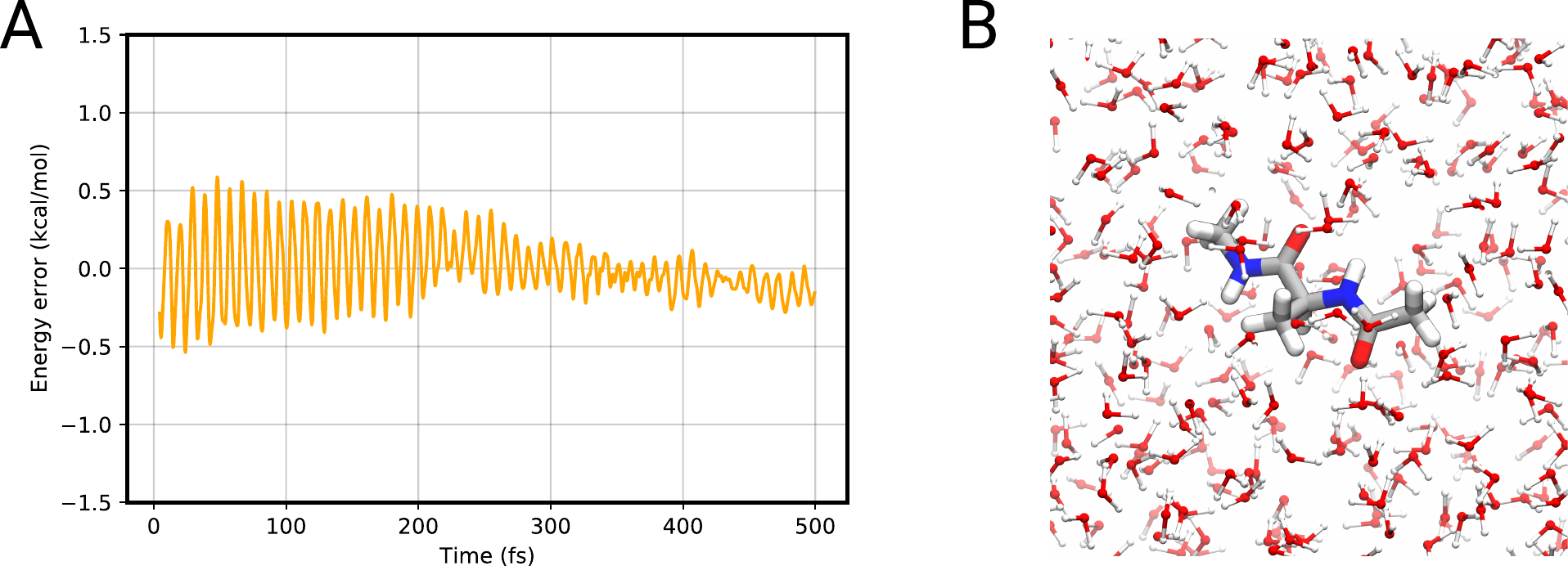}
     \caption{Results of the MD test performed on of alanine dipeptide. (A) Energy change during the MD simulation. (B) Representation of the simulated system.}
     \label{fig:optimization}
 \end{figure}

One of the most important strengths of the herein presented library is its ability to compute the entire potential (\textit{i.e.} including bonded and VdW terms) and its geometrical gradients as this allows performing geometry optimizations and MD simulations. Our OpenMMPol-PySCF interface is able to perform both those types of calculation. To demonstrate this capability we report here a short MD on a model system.
The chosen system is alanine dipeptide solvated with explicit water (see  Figure~\ref{fig:optimization}B), which has 9876 polarizable atoms (alanine dipeptide and water molecules) and a QM subsystem of 22 atoms. The MM part is described with AMOEBA while the QM part with B3LYP/6-31G (114 basis functions). The MD was propagated with the internal integrator available in PySCF. 
Figure~\ref{fig:optimization}A shows that the energy is conserved during the MD simulation, and no drift is observed, meaning that the energy and the gradients are well consistent.

\begin{figure}[!ht]
     \centering
     \includegraphics[scale=0.4]{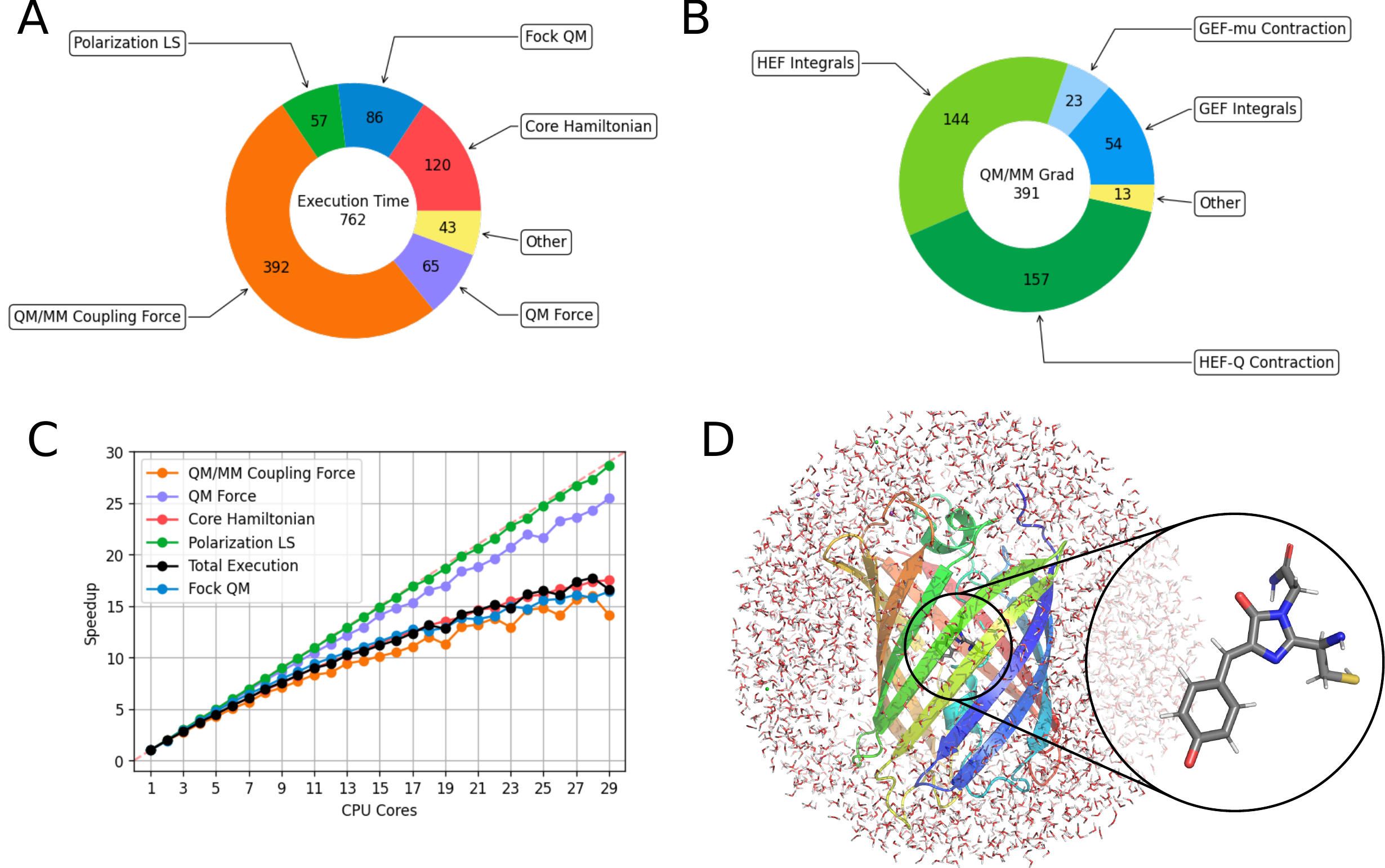}
     \caption{(A) Timing for a complete SCF energy and gradients calculation performed with PySCF interfaced with OpenMMPol. To keep the diagram clear, all the steps that are below 5\% of total execution are neglected here, and summed under ``Other". Times are expressed in seconds. (B) Timing for the calculation of the derivatives QM/MM couplings (see Eq.~\ref{eq:der_e_qmmm} and Eq.~\ref{eq:grd_wrt_qm_eqmmm}). (C) Speedup vs number of CPU curves for the different steps of the SCF calculation analyzed in panel A, and for the total calculation time (in black). (D) Graphical representation of the solvated DRONPA protein used for the benchmark; the QM region (excluding the link atoms) is represented in the zoom.}
     \label{fig:SCFtime}
 \end{figure}

To investigate the computational efficiency and bottlenecks of the interfaced code, we have performed and profiled a full SCF energy and gradients calculation on a system composed by 14578 polarizable atoms, and a QM part with 466 basis functions (see Figure~\ref{fig:SCFtime}C). The system used in this test is the DRONPA Green Fluorescent Protein\cite{Mizuno2008}, solvated in explicit water, as already prepared in a previous paper by our group.\cite{Nottoli20213l}

From Figure \ref{fig:SCFtime}A it is clear that the main bottlenecks in the current implementation of the SCF is in the solution of the linear system and in the calculation of the core Hamiltonian. This is something completely expected; the first operation, performed on a pretty large system as the one used in this example, suffers from the non-optimal $\mathcal O(N^2)$ scaling in computing electrostatic interaction already underlined in the first part of this paragraph. The calculation of core Hamiltonian is done contracting atomic orbital integrals with the static charges (and also dipoles and quadrupoles, for AMOEBA) and is basically limited by the computation of the integrals themselves. A substantial speedup in the SCF cycle was also attained by storing (when it is possible) the electric field integrals in memory, neglecting all the elements close to zero (this, especially for large molecules/basis sets, saves a considerable amount of memory). In this way the calculation of the QM electric field, needed to assemble the RHS of eq. \ref{eq:pol_eq} and the contribution to the Fock matrix (eq. \ref{eq:veff_mmpol}) which are performed on each SCF cycle do not require the calculation of the large array of atomic orbital integrals.

Almost half of the calculation time is dedicated to computing forces on QM and MM atoms, again the calculation of an even larger number (with respect to SCF core Hamiltonian) of integrals is the main bottleneck (see Fig.~\ref{fig:SCFtime}B), together with their contraction with multipoles.
In the evaluation of those results, it should be considered that the integrals for computing high order derivatives of the electric field are written with a code generator, and are likely not fully optimized. As a matter of fact, while in the SCF the largest overhead is actually due to the polarization (namely the solution of the linear system), for the forces, the overhead is due to the large number of static dipoles and quadrupoles. 
For an AMBER-like formulation, where only fixed charges are needed, the overall cost for the same system would be greatly reduced.

 In Figure \ref{fig:SCFtime}C the parallel performance of the whole SCF are analyzed. Overall we notice that, while the scaling is not optimal, it is still good. The parts of the code that exhibit a lower scaling are the calculations of QM/AMOEBA interactions which, as pointed out before, are basically calculation and contraction of atomic integrals with density matrices or electrostatic multipoles. Since the contraction operation is performed with the highly parallel optimized \texttt{dgemm} routine from BLAS library, we can speculate that the suboptimal scaling observed in those part of the code can be ascribed to the integral calculations which is expected to be suboptimal for the reasons mentioned above.

\section{Summary}

We have presented the OpenMMPol library that can be used in combination with most QM codes to run polarizable QM/MM calculations of energies, properties and dynamics of complex systems. 
OpenMMPol is the first library implementation which is able to compute energies and gradients also including the classical terms of the potential. Despite the low computational cost of those terms, this feature should not be overlooked, as it allows to perform geometry optimization and MD directly from the QM code (or with a MD driver) without any interface to external programs to compute the classical interactions. Even though our focus is mainly on polarizable QM/MM, we underline that also non-polarizable environments are completely supported, and therefore once the interface is in operation it could also be used to perform e.g. electrostatic-embedding QM/AMBER calculations. 

The OpenMMPol library is released with a permissive FOSS license and is completely agnostic with respect to the specific QM software and methods. The modern build system adopted, and the overall clear and simple programming style of the source code, and its flexibility, make it easy to use this library as a platform for further development of the method and experimentation of new algorithms.
As a demonstration, and as a tutorial-code, we have presented an interface with the popular PySCF code for performing QM/AMOEBA calculations at HF, DFT (with gradients) and TD-DFT levels.
The library will also allow to expand the number of methods interfaced with powerful polarizable QM/MM environment. In particular, we are currently working on interfacing the library to multiconfigurational SCF codes, in order to explore non-adiabatic dynamics methods.

As a final remark we want to underline that a great attention was paid to the computational efficiency and the parallel scalability of our code. To this end the main limiting factor is the quadratically scaling electrostatics which makes calculations with more than 100k atoms unpractical. 
To extend even further the field of application, in the near future a new version of the library implementing FMM methods for solving the electrostatics will be released.

\section*{Supplementary Material}
Derivation of integrals needed for QM-AMOEBA and QM-MMPol coupling and derivative in term of 3-centers-2-electron integrals (also expressed as \texttt{libcint} functions); Description of \texttt{.mmp} legacy format.

\section*{Acknowledgements}
All authors acknowledge financial support from ICSC-Centro Nazionale di Ricerca in High Performance Computing, Big Data, and Quantum Computing, funded by the European Union -- Next Generation EU -- PNRR, Missione 4 Componente 2 Investimento 1.4.
M.B. and B.M. gratefully acknowledge funding from the European Research Council under grant ERC-AdG-786714 (LIFETimeS).
M.B. acknowledges the CINECA award under the ISCRA initiative, for the availability of high-performance computing resources and support (OMMPPOO).

\section*{Data availability}
The input files for the systems used for the tests and benchmarks presented in the present paper are provided together with the OpenMMPol code in the \texttt{tests} directory of the git repository.

\section*{Code availability}
OpenMMPol is released on GitHub and can be downloaded from https://github.com/Molecolab-Pisa/OpenMMPol; the version used in the present paper is 1.0.0.

The interface with PySCF is also provided on GitHub and can be downloaded from https://github.com/Molecolab-Pisa/pyscf-openmmpol (commit \texttt{fb2a75f}).

\bibliography{new_biblio}

\end{document}


\author{Mattia Bondanza}
\affiliation{Dipartimento di Chimica e Chimica Industriale, University of Pisa, via G. Moruzzi 13, 56124, Pisa, Italy}
\author{Tommaso Nottoli}
\affiliation{Dipartimento di Chimica e Chimica Industriale, University of Pisa, via G. Moruzzi 13, 56124, Pisa, Italy}
\author{Michele Nottoli}
\affiliation{Institute of Applied Analysis and Numerical Simulation, Universit\"at Stuttgart, Pfaffenwaldring 57, D-70569, Stuttgart, Germany}
\author{Lorenzo Cupellini}
\affiliation{Dipartimento di Chimica e Chimica Industriale, University of Pisa, via G. Moruzzi 13, 56124, Pisa, Italy}
\author{Filippo Lipparini}
\affiliation{Dipartimento di Chimica e Chimica Industriale, University of Pisa, via G. Moruzzi 13, 56124, Pisa, Italy}
\author{Benedetta Mennucci}
\email{benedetta.mennucci@unipi.it}
\affiliation{Dipartimento di Chimica e Chimica Industriale, University of Pisa, via G. Moruzzi 13, 56124, Pisa, Italy}

\title{Supplementary Material for ``The OpenMMPol Library for Polarizable QM/MM Calculations of Properties and Dynamics''}

\maketitle

\section{Electrostatic Integrals for MMPol and AMOEBA}
In modern QM software based on localized Gaussian basis function, electrostatic integrals are computed as three-center two-electron (3C2E) integrals according to the following equivalence:
\begin{align}
    \langle \mu(\vec{r}_1) | \hat{V}(\vec{R}) | \nu(\vec{r}_1) \rangle &=
    \langle \mu(\vec{r}_1) | \frac{1}{|\vec{R} - \hat{\vec{r}}_1|} | \nu(\vec{r}_1) \rangle \\
    &= \langle \mu(\vec{r}_1) \nu(\vec{r}_1) | \frac{1}{|\hat{\vec{r}}_2 - \hat{\vec{r}}_1|} | \delta_{\vec{R}}(\vec{r}_2) \rangle,
\end{align}
where $\delta$ is a narrow s-type Gaussian orbital (mimicking a Kroneker delta in 3-dimension) centered in $\vec{R}$.
This apparently strange approach is used because of its computational efficiency.

When is needed to compute derivatives of electrostatic potential, the same trick can be used; for example to compute the electric field integrals:
\begin{align}
    \langle \mu(\vec{r}_1) | \hat{E}(\vec{R}) | \nu(\vec{r}_1) \rangle &=
    \langle \mu(\vec{r}_1) | \nabla_{\vec{r}_1} \frac{1}{|\vec{R} - \hat{\vec{r}}_1|} | \nu(\vec{r}_1) \rangle \\
    &= \langle \nabla_{\vec{r}_1}  \mu(\vec{r}_1) | \frac{1}{|\vec{R} - \hat{\vec{r}}_1|} | \nu(\vec{r}_1) \rangle + 
    \langle \mu(\vec{r}_1) | \frac{1}{|\vec{R} - \hat{\vec{r}}_1|} |  \nabla_{\vec{r}_1} \nu(\vec{r}_1) \rangle \\
    \label{eq:3c2e_field}
    &= \langle  \nabla_{\vec{r}_1} \mu(\vec{r}_1) \nu(\vec{r}_1) | \frac{1}{|\hat{\vec{r}}_2 - \hat{\vec{r}}_1|} | \delta_{\vec{R}}(\vec{r}_2) \rangle + \langle   \mu(\vec{r}_1) \nabla_{\vec{r}_1} \nu(\vec{r}_1) | \frac{1}{|\hat{\vec{r}}_2 - \hat{\vec{r}}_1|} | \delta_{\vec{R}}(\vec{r}_2) \rangle;
\end{align}
note that, if the 3C2E integrals in eq. \ref{eq:3c2e_field} are computed on the whole basis set at once (thus creating a matrix $N_\mathrm{basis} \times N_\mathrm{basis}$) as it is normally done, the two terms are just the transposed elements of the same integral matrix:
\begin{align}
    \mathbf{E}_{\mu,\nu}(\vec{R}) &= \langle \mu(\vec{r}_1) | \hat{E}(\vec{R}) | \nu(\vec{r}_1) \rangle\\
    \mathbf{I}_{\mu,\nu}(\vec{R}) &= \langle  \nabla_{\vec{r}_1} \mu(\vec{r}_1) \nu(\vec{r}_1) | \frac{1}{|\hat{\vec{r}}_2 - \hat{\vec{r}}_1|} | \delta_{\vec{R}}(\vec{r}_2) \rangle\\
    \mathbf{E} &= \mathbf{I} + \mathbf{I}^\dagger.
\end{align}

In the following we will use the dagger after a 3C2E braket to indicate the exchange of the two nuclear basis function:
\begin{equation}
    \langle  \hat{\mathcal{O}}_1 \mu(\vec{r}_1) \hat{\mathcal{O}}_2 \nu(\vec{r}_1) | \hat{\mathcal{O}} | \delta(\vec{r}_2) \rangle ^\dagger = \langle  \hat{\mathcal{O}}_1 \nu(\vec{r}_1) \hat{\mathcal{O}}_2 \mu(\vec{r}_1) | \hat{\mathcal{O}} | \delta(\vec{r}_2) \rangle.
\end{equation}

In the following table we will provide a list of all the electrostatic integrals needed for MMPol and/or AMOEBA SCF calculations (included gradients) and the \textit{recipes} to build them from 3C2E integrals; for further clarity we also provide the integral names in \texttt{libcint} format for 3C2E integrals. The operator $\frac{1}{|\hat{\vec{r}}_2 - \hat{\vec{r}}_1|}$ will be denoted as $\hat{R}_{12}$ for brevity and the explicit dependence from electronic coordinates will also be neglected.
\begin{table}[ht]
\caption{Expression of AMOEBA and MMPol integrals in terms of 3C2E integrals.}
\centering
\begin{tabular}{ c| c }
    Atomic Orbital Integral & 3C2E Expansion\\
    \hline
     $\langle \mu | \hat{V}(\vec{R}) | \nu\rangle$ & 
     $\langle \mu \nu | \hat{R}_{12} | \delta \rangle$  \\
     
     $\langle \mu | \hat{\vec{E}}(\vec{R}) | \nu\rangle$ & 
     $\langle \nabla \mu \nu | \hat{R}_{12} | \delta \rangle + 
     \langle \nabla \mu \nu | \hat{R}_{12} | \delta \rangle^\dagger$ \\
     
     $\langle \mu | \hat{\nabla \vec{E}}(\vec{R}) | \nu\rangle$ &
     $\langle \nabla^2 \mu \nu | \hat{R}_{12} | \delta \rangle + 
     \langle \nabla^2 \mu \nu | \hat{R}_{12} | \delta \rangle^\dagger + 
     2 \langle \nabla \mu \nabla \nu | \hat{R}_{12} | \delta \rangle$ \\

     $\langle \mu | \hat{\mathbf{H}}_{\vec{E}}(\vec{R}) | \nu\rangle$ &
     $\langle \nabla^3 \mu \nu | \hat{R}_{12} | \delta \rangle + 
     \langle \nabla^3 \mu \nu | \hat{R}_{12} | \delta \rangle^\dagger + 
     3\Bigg( \langle \nabla^2 \mu \nabla \nu | \hat{R}_{12} | \delta \rangle +
     \langle \nabla^2 \mu \nabla \nu | \hat{R}_{12} | \delta \rangle^\dagger \Bigg)$ \\

     $\langle \nabla \mu | \hat{V}(\vec{R}) | \nu \rangle$ & 
     $\langle \nabla \mu \nu | \hat{R}_{12} | \delta \rangle$ \\

    $\langle \nabla \mu | \hat{E}(\vec{R}) | \nu \rangle$ & 
    $\langle \nabla^2 \mu \nu | \hat{R}_{12} | \delta \rangle + 
     \langle \nabla \mu \nabla \nu | \hat{R}_{12} | \delta \rangle$ \\

    $\langle \nabla \mu |  \hat{\nabla \vec{E}}(\vec{R}) | \nu \rangle$ & 
    $\langle \nabla^3 \mu \nu | \hat{R}_{12} | \delta \rangle + 
     2 \langle \nabla^2 \mu \nabla \nu | \hat{R}_{12} | \delta \rangle +
     \langle \nabla^2 \mu \nabla \nu | \hat{R}_{12} | \delta \rangle^\dagger$ \\

\end{tabular}
\end{table}

\begin{table}[ht]
\caption{Equivalence between 3C2E integrals and routine names in libcint.}
\centering
\begin{tabular}{ c| c }
    3C2E Integral & Name in \texttt{libcint} \\
    \hline
     $\langle \mu \nu | \hat{R}_{12} | \delta \rangle$ & 
     \texttt{INT3C2E} \\
     
     $\langle \nabla \mu \nu | \hat{R}_{12} | \delta \rangle$  & 
     \texttt{INT3C2E\_IP1}\\
     
     $\langle \nabla^2 \mu \nu | \hat{R}_{12} | \delta \rangle$ &
     \texttt{INT3C2E\_IPIP1} \\
     
     $\langle \nabla \mu \nabla \nu | \hat{R}_{12} | \delta \rangle$ & 
     \texttt{INT3C2E\_IPVIP1} \\
     
     $\langle \nabla^3 \mu \nu | \hat{R}_{12} | \delta \rangle$ &
     \texttt{INT3C2E\_IPIPIP1} \\
    
     $\langle \nabla^2 \mu \nabla \nu | \hat{R}_{12} | \delta \rangle$ &
     \texttt{INT3C2E\_IPIPVIP1}
\end{tabular}
\end{table}

\section{.mmp File Format}
The .mmp file is a file only used internally during previous phases of development of MMPol model, it is implemented in the library only for backward compatibility reasons and it is not recommended to be used in new applications. 
It is a formatted file in free format, with some differences between version 2 and 3 of the format; each line has a specific meaning according to the following table:

\begin{table}[ht]
\caption{Specification of .mmp format}
\centering
\begin{longtable}{ c | c | l | c}
    lines & Type & Description & Notes \\
    \hline
    1 & integer & Version number (2 or 3 allowed) &  \\
    2 & integer & Job Type \footnote{0=single point, 1=tinker QM/MM, 2=electronic energy transfer} & Ignore\\
    3 & integer & Verbosity Flag & Ignored\\
    4 & integer & Force Field Type \footnote{0=Amber-like, 1=AMOEBA-like} & \\
    5 & integer & Force Field Sub-Type \footnote{0=Amber Wang AL or AMOEBA, 1=Amber Wang DL, 2=Amber Thole}& Ignored \\
    6 & integer & Disabled flag & Ignored \\
    7 & float   & Dipole-dipol damping & Ignored \\
    8 & integer & Method for linear system\footnote{0=default, 1=matrix inversion, 2=DIIS 3=conjugate gradients} & Ignored \\
    9 & integer & Method for matrix-vector\footnote{0=default, 1=in-memory, 2=on-the-fly double loop, 3=on-the-fly FMM, 4=on-the-fly FMM forced} & Ignored \\
    10 & integer & Convergence for linear system\footnote{$t=10{-n}$} & Ignored \\
    11 & integer & Accuracy for FMM\footnote{-1: $10^{-5}$ RMS and $10^{-4}$ max; 0: $10^{-6}$ RMS and $10^{-5}$ max; 1: $10^{-7}$ RMS and $10^{-6}$ max; 2: maximum accuracy.} & Ignored \\
    12 & float   & FMM box size (\AA) & Ignored \\
    13 & float   & FMM box size (\AA) for MMPol-PCM & Ignored, only v.3 \\
    14 & integer & Continuum solvation\footnote{0=none, 1=ddCOSMO, 2=ddPCM} & Ignored \\
    15 & integer & Maximum angular momentum for dd & Ignored \\
    16 & integer & Number of Lebedev points for dd & Ignored \\
    17 & integer & Convergence for ddPCM\footnote{$t=10{-n}$} & Ignored \\ 
    18 & float   & Dielectric constant for continuum solvent & Ignored \\
    19 & float   & Optical dielectic constant for continuum solvent & Ignored, only v.3 \\
    20 & float   & Relative size of switching region for PCM & Ignored \\
    21 & integer & Cavity for PCM\footnote{0=default, 1=SAS with UFF radii plus probe radius, 2=Cavity from input file, 3=VdW cacity using Bondi's radi scaled by 1.1, 4=SAS cavity using Bondi's radii plus probe radius.} & Ignored \\
    22 & float   & Probe radius & Ignored \\
    23 & integer & Number of MM atoms & \\
    24 & integer & Number of cavity's spheres & Ignored \\
    25 (N\footnote{\label{nmm}Number of MM atoms} lines)& integer & Atomic numbers & \\
    26 (N\footref{nmm} lines) & 3 float & Coordinates (\AA) & \\
    27 (N\footref{nmm} lines) & integer & Residue number & Ignored \\
    28 (N\footref{nmm} lines) & 1 or 10 float & Fixed multipoles & \\
    29 (N\footref{nmm} lines) & float & Isotropic polarizabilities & \\
    30 (N\footref{nmm} lines) & 8 integer & Indexes of connecte atoms\footnote{1-based, 0 means no connection} & \\
    31 (N\footref{nmm} lines) & 120 integer & Index of atoms in the same polgroup\footnote{1-based, 0 means no connection} & For AMOEBA only \\
    32 (N\footref{nmm} lines) & 4 integer & Rotation conventions\footnote{0=do not rotate, 1=z-then-x, 2=bisector, 3=z-only, 4=z-bisector, 5=3-fold} and reference atoms\footnote{1-based, 0 means not used} & For AMOEBA only \\

\end{longtable}
\end{table}